\documentclass{emulateapjx}

\def\al{\alpha}

\def\de{\delta}
\def\ep{\epsilon}

\def\ze{\zeta}

\def\ka{\kappa}
\def\la{\lambda}

\def\si{\sigma}
\def\vs{\varsigma}

\def\ps{\psi}
\def\om{\omega}

\def\De{\Delta}

\def\La{\Lambda}
\def\Si{\Sigma}

\def\Om{\Omega}

\def\cl{{\cal L}}

\def\Frac#1#2{{\textstyle{{#1}\over {#2}}}}
\def\lsim{\mathrel{\rlap{\lower4pt\hbox{\hskip1pt$\sim$}}
    \raise1pt\hbox{$<$}}}
\def\gsim{\mathrel{\rlap{\lower4pt\hbox{\hskip1pt$\sim$}}
    \raise1pt\hbox{$>$}}}

\def\prt{\partial}

\newcommand{\beq}{\begin{equation}}
\newcommand{\eeq}{\end{equation}}
\newcommand{\bea}{\begin{eqnarray}}
\newcommand{\eea}{\end{eqnarray}}
\newcommand{\rf}[1]{(\ref{#1})}

\def\mbf#1{\mbox{\boldmath$#1$}}
\def\syjm#1#2{\phantom{}_{#1}Y_{#2}}

\def\kf{\hat k_F}
\def\kaf{\hat k_{AF}}
\def\kfd#1{k_{F}^{(#1)}}
\def\kafd#1{k_{AF}^{(#1)}}

\def\kjm#1#2#3{k^{(#1)}_{(#2)#3}}
\def\kI{\kjm{d}{I}{jm}}
\def\kE{\kjm{d}{E}{jm}}
\def\kB{\kjm{d}{B}{jm}}
\def\kV{\kjm{d}{V}{jm}}
\def\kVt{\kjm{3}{V}{jm}}
\def\kIdjm#1#2{\kjm{#1}{I}{#2}}

\def\kVdjm#1#2{\kjm{#1}{V}{#2}}

\def\etal{et al.}

\begin{document}

\title{Astrophysical Tests of Lorentz and CPT Violation with Photons}

\author{V.\ Alan Kosteleck\'y\altaffilmark{1} and Matthew Mewes\altaffilmark{2}}

\altaffiltext{1}{Physics Department, Indiana University, 
Bloomington, IN 47405, U.S.A.}
\altaffiltext{2}{Physics Department, Marquette University,
Milwaukee, WI 53201, U.S.A.}

\begin{abstract}
A general framework for tests of Lorentz invariance 
with electromagnetic waves is presented,
allowing for operators of arbitrary mass dimension. 
Signatures of Lorentz violations include
vacuum birefringence, vacuum dispersion, and anisotropies.
Sensitive searches for violations
using sources such as active galaxies, gamma-ray bursts,
and the cosmic microwave background are discussed.
Direction-dependent dispersion constraints are obtained 
on operators of dimension 6 and 8 using
gamma-ray bursts and the blazar Markarian 501.
Stringent constraints on operators of dimension 3
are found using 5-year data from 
the Wilkinson Microwave Anisotropy Probe.
No evidence appears for isotropic Lorentz violation,
while some support at 1$\si$ is found
for anisotropic violation.
\end{abstract}

\keywords{ relativity --- gravitation --- 
cosmic microwave background --- gamma rays: bursts --- 
galaxies: active}

\maketitle

Recent years have seen a resurgence in tests of relativity,
spurred in part by the prospect of relativity violations
arising in a unified description of nature
\citep{ks89,kp91}.
Experimental searches for violations of Lorentz invariance,
the symmetry underlying relativity,
have been performed in a wide range of systems
(for data tables, see \citet{tables}).
Historically,
experiments probing the behavior of light have been central 
in confirming relativity.
Contemporary versions 
of the classic Michelson-Morley and Kennedy-Thorndike experiments
\citep{cavities1,cavities2,cavities4}
remain among the most sensitive tests today.

Some tight constraints on relativity violations
have been achieved by seeking tiny changes in light 
that has propagated over astrophysical distances.
Many of these search for a change in polarization
resulting from vacuum birefringence,
using sources such as galaxies 
\citep{cfj,sme2,km_agn,km},
gamma-ray bursts (GRB)
\citep{grb_bire1,grb_bire2,km_grb,grb_bire3,grb_bire4},
and the cosmic microwave background (CMB)
\citep{cmb_bire12,cmb_bire23,km_cmb,cmb_bire13,wmap_cpt,xia,kdm}.
Others seek a frequency-dependent velocity
arising from vacuum dispersion,
using GRB, pulsars, and blazars 
\citep{disp,km,boggs,mmtp,emnss,lps,magic}.
Here,
we present a general theoretical framework 
that characterizes Lorentz-violating effects 
on the vacuum propagation of electromagnetic waves
and includes operators of all mass dimensions.
We discuss several techniques that
can be used to search for the unconventional signals 
of Lorentz violation,
Using vacuum-dispersion constraints
from GRB and the blazar Markarian 501,
we place new direction-dependent limits
on several combinations of coefficients for Lorentz violation.
We also perform a search for Lorentz violations
in the 5-year results from  
the Wilkinson Microwave Anisotropy Probe (WMAP)
\citep{wmap_cpt,wmap5yr1,wmap5yr2},
finding some evidence for anisotropic violations
but no support for isotropic violations.

At attainable energies,
violations of Lorentz invariance are described
by a framework called 
the Standard-Model Extension (SME)
\citep{sme1,sme2,sme3}
that is based on effective field theory
\citep{kp95}.
Approaches outside field theory also exist
\citep{review}.
The SME characterizes 
all realistic violations affecting known particles and fields,
while incorporating otherwise established physics.
Much of the work on Lorentz violation
has focused on the minimal SME,
which restricts attention to 
gauge-invariant operators of renormalizable dimension.
In this work,
we consider the gauge-invariant pure-photon sector of the full SME
with Lorentz-violating operators of arbitrary dimension,
which has Lagrange density 
\citep{km_cmb}
\bea
\cl &=&  -\Frac 1 4 F_{\mu\nu}F^{\mu\nu}
+\Frac 1 2 \ep^{\ka\la\mu\nu}A_\la (\kaf)_\ka F_{\mu\nu}
\nonumber \\
&& \quad 
- \Frac 1 4 F_{\ka\la} (\kf)^{\ka\la\mu\nu} F_{\mu\nu}\ ,
\label{lagrangian}
\eea
where $A_\mu$ is the 4-potential with field strength $F_{\mu\nu}$.
In a flat background with energy-momentum conservation,
the Lorentz violation arises through the differential operators
\bea
(\kaf)_\ka &=&
\hspace{-3pt}\sum_{d=\mbox{\scriptsize odd}} 
{(\kafd{d})_\ka}^{\al_1\ldots\al_{(d-3)}} 
\prt_{\al_1}\ldots\prt_{\al_{(d-3)}} ,
\label{kafs}\\
(\kf)^{\ka\la\mu\nu} &=& 
\hspace{-5pt}\sum_{d=\mbox{\scriptsize even}} 
(\kfd{d})^{\ka\la\mu\nu\al_1\ldots\al_{(d-4)}} 
\prt_{\al_1}\ldots\prt_{\al_{(d-4)}} .
\label{kfs}
\eea
The constant coefficients 
${(\kafd{d})_\ka}^{\al_1\ldots\al_{(d-3)}}$
and
$(\kfd{d})^{\ka\la\mu\nu\al_1\ldots\al_{(d-4)}}$
characterize the degree of Lorentz violation.
The former control CPT-odd operators 
and are nonzero for odd dimension $d\geq 3$,
while the latter control CPT-even operators 
and are restricted to even $d\geq 4$.

The Lagrange density \rf{lagrangian}
yields modified Maxwell equations.
At leading order in coefficients for Lorentz violation,
two plane-wave solutions exist.
The corresponding two modified dispersion relations
can be written in the form
\beq
p(\om) \approx [1 + \vs^0 \mp
\sqrt{(\vs^1)^2+(\vs^2)^2+(\vs^3)^2}\, ] \om ,
\label{disp}
\eeq
where $p$ and $\om$ are the wavenumber and frequency,
respectively.
It follows that electromagnetic waves 
generically contain two propagating modes
with different velocities and polarizations.
The symbols
$\vs^0$,
$\vs^1$,
$\vs^2$, and
$\vs^3$ 
represent certain combinations of coefficients
for Lorentz violation,
and they depend on the frequency $\om$
and direction of propagation $\mbf{\hat p}$.
With convenient normalizations,
$\vs^1$, $\vs^2$, and $\vs^3$
are the Stokes parameters
$s^1=Q$, $s^2=U$, and $s^3=V$
of the faster mode,
while $\vs^0$
is a scalar combination analogous to the intensity $s^0=I$.
These four combinations completely control 
the leading-order effects of Lorentz violation 
on light propagating through empty space.
The combination $\vs^3$ depends only 
on the coefficients
${(\kafd{d})_\ka}^{\al_1\ldots\al_{(d-3)}}$,
while $\vs^0$, $\vs^1$, and $\vs^2$
depend only on the coefficients 
$(\kfd{d})^{\ka\la\mu\nu\al_1\ldots\al_{(d-4)}}$.

It is convenient to identify a minimal set 
of coefficient combinations 
that affect light propagating \it in vacuo. \rm
This can be accomplished through 
spherical-harmonic decomposition.
Since $\vs^0$, $\vs^3$ are rotation scalars
while $\vs^1$, $\vs^2$ are rotation tensors,
their decomposition must involve some form of
tensor spherical harmonics.
The spin-weighted harmonics
$\syjm{s}{jm}(\mbf{\hat p})$
provide a well-understood set 
\citep{sYjm1,sYjm2}.
The index $s$ is the spin weight,
which up to a sign is equivalent to helicity.
Decomposing yields 
\bea
\vs^0 &=& 
\sum_{djm} \om^{d-4} \, \syjm{0}{jm}(\mbf{\hat n})\,  \kI \ ,
\nonumber \\
\vs^1\pm i \vs^2 &=&
\sum_{djm} \om^{d-4} \, \syjm{\pm2}{jm}(\mbf{\hat n})\,
\big(\kE\mp i\kB\big) \ ,
\nonumber \\
\vs^3 &=&
\sum_{djm} \om^{d-4} \, \syjm{0}{jm}(\mbf{\hat n})\, \kV \ ,
\label{vac_exp}
\eea
where $j\leq d-2$ and $\mbf{\hat n}=-\mbf{\hat p}$ is a unit vector
pointing to the source in astrophysics tests.

With this decomposition,
all types of Lorentz violations 
for propagation {\it in vacuo}
can now be simply characterized 
using four sets of spherical coefficients,
$\kI$, $\kE$, $\kB$ for CPT-even effects
and $\kV$ for CPT-odd effects.
For each coefficient,
the underlying Lorentz-violating operator
has mass dimension $d$
and eigenvalues of total angular momentum given by $jm$, 
as usual.
For light from astrophysical sources,
dispersion arises when the speed of
propagation depends on frequency,
which occurs for any nonzero coefficient with $d\neq 4$.
Birefringence results when the usual degeneracy
among polarizations is broken,
for which at least one of $\kE$, $\kB$, $\kV$
is nonzero.
For example,
all operators producing lightspeed corrections 
that are linear in the energy
have $d=5$ and are necessarily birefringent.
The only coefficients for nonbirefringent dispersion
are therefore
$\kI$ with even $d\geq 6$.
Since birefringence tests using polarimetry are typically 
many orders of magnitude more sensitive 
than dispersion tests using timing,
in the following discussion of dispersion 
we focus only on coefficients for nonbirefringent dispersion. 

Tests for vacuum dispersion seek differences 
in the velocity of light at different wavelengths.
In the present context
with zero birefringent coefficients,
the change in velocity is $\de v \simeq -\vs^0$.
We see from Eq.\ \rf{vac_exp}
that the velocity generically depends 
on the direction $\mbf{\hat n}$ as well as the frequency $\om$.
Typical analyses study explosive or pulsed sources of radiation 
producing light over a wide wavelength range in short time periods,
comparing the arrival times of different wavelengths.
This idea has been the focus of many searches
based on modified dispersion relations
\citep{disp,km,boggs,mmtp,emnss,lps,magic}.
Many of these studies assume isotropic violations,
which corresponds to the limit $j=m=0$.
However,
at each dimension $d$,
this isotropic restriction misses $(d^2 - 2d - 2)$ possible effects
from anisotropic violations.

To calculate arrival-time differences
in an expanding universe,
some care is required 
\citep{jacob}.
In the present case,
the photons propagate between two comoving objects, 
so the relevant coordinate interval is
$dl_c = (1+z) dl_p = -v_z dz / H_z$.
Here,
$v_z$ is the particle velocity at redshift $z$,
and 
$H_z = H_0(\Om_r\ze^4 + \Om_m\ze^3
+\Om_k\ze^2+\Om_\La)^{1/2}$
with $\ze = 1+z$
is the Hubble expansion rate at $z$ in terms of the
present-day Hubble constant $H_0 \simeq71$ km/s/Mpc,
radiation density $\Om_r\simeq 0$,
matter density $\Om_m\simeq 0.27$,
vacuum density $\Om_\La\simeq 0.73$, 
and curvature density 
$\Om_k = 1-\Om_r-\Om_m-\Om_\La$.
The total coordinate distance is the same for all wavelengths,
but the travel times may differ.
Integrating $dl_c$ from the same initial
time to the two arrival times for the two velocities
gives a relation for the arrival-time difference $\De t$,
which depends on the two energies 
and the source location on the sky.
For the present case with Lorentz violation at dimension $d$,
we find
\beq
\De t \approx -\De \om^{d-4} 
\int_0^z \frac{(1+z)^{d-4}}{H_z}dz
\sum_{jm}
\syjm{0}{jm}(\mbf{\hat n}) \kI ,
\label{det}
\eeq
where $\De \om^{d-4}$
is the difference in $\om^{d-4}$
between the two frequencies.

As an illustration,
consider the bright gamma-ray burst GRB 021206
at right ascension $240^\circ$ and 
declination $-9.7^\circ$.
Over energies from 3 to 17 MeV,
arrival-time differences are no more than $\De t < 4.8$ ms 
for this source at $z \simeq 0.3$
\citep{boggs}.
Numerical integration of Eq.\ \rf{det}
leads to a bound on one direction-specific combination 
of the 25 independent coefficients
for nonbirefringent dispersion with $d=6$:
\beq
\sum_{jm}\syjm{0}{jm}(99.7^\circ,240^\circ)\,  
\kIdjm{6}{jm} < 1\times 10^{-16} \mbox{ GeV}^{-2}\ .
\label{d6}
\eeq
For the 49 independent nonbirefringent dispersive operators
with $d=8$,
we obtain
\beq
\sum_{jm}\syjm{0}{jm}(99.7^\circ,240^\circ)\,  
\kIdjm{8}{jm} < 3\times 10^{-13} \mbox{ GeV}^{-4}\ .
\label{d8}
\eeq
Operators with higher $d$ can be treated similarly.
Note that many sources are required to constrain fully
the coefficient space for a given $d$.
In contrast,
only one source is needed to constrain fully 
the corresponding coefficient 
in the restrictive isotropic limit $j=m=0$.
In this limit,
the bounds \rf{d6} and \rf{d8} reduce to
$\kIdjm{6}{00} < 4\times 10^{-16} \mbox{ GeV}^{-2}$
and
$\kIdjm{8}{00} < 9\times 10^{-13} \mbox{ GeV}^{-4}$,
respectively.

As another example,
consider Markarian 501,
which lies at $z\simeq 0.03$.
This source produces flares 
with photon energies in the TeV range,
making it particularly sensitive
to an energy-dependent velocity
and also to threshold analyses
\citep{threshold}.
A recent analysis of observations 
by the MAGIC collaboration 
found some evidence for a nonbirefringent velocity defect 
of the form 
$\de v = - \om/M$ or $\de v = - \om^2/M^2$
\citep{magic}.
The first case is incompatible with the present treatment;
a reanalysis incorporating the necessary birefringence 
could yield comparatively weak but compatible new bounds. 
The second case suggests dispersion 
with $M\simeq 6^{+5}_{-1}\times 10^{10}$ GeV,
assuming an arrival-time lag 
due entirely to nonbirefringent Lorentz violation.
For $d=6$, 
this yields the single constraint 
\beq
\sum_{jm}\syjm{0}{jm}(50.2^\circ,253^\circ)\,  
\kIdjm{6}{jm} \simeq  3^{+1}_{-2}\times 10^{-22} \mbox{ GeV}^{-2}\ ,
\eeq
consistent with the GRB bound \rf{d6}.
In the isotropic limit,
this becomes
$\kIdjm{6}{00} \simeq 10^{+4}_{-7}\times 10^{-22} \mbox{ GeV}^{-2}$.

Next,
we consider tests for vacuum birefringence.
In birefringent scenarios,
the two plane-wave eigenmodes travel 
at slightly different velocities,
which alters their superposition 
and hence the net polarization of the light 
as it propagates in free space.
The polarization change is equivalent to 
a rotation of the Stokes vector
$\mbf s = (s^1,s^2,s^3)^T$
about the birefringent axis 
$\mbf\vs = (\vs^1,\vs^2,\vs^3)^T$.
The total rotation angle is equal to 
the relative phase change between the two eigenmodes.
Infinitesimally,
the rate of rotation is
$d\mbf s/dt =  -i\Si\cdot\mbf s$,
where $\Si^{ab} = -2i\om\ep^{abc}\vs^c$ 
is the rotation generator.
Integration from source redshift $z$ to $0$
taking into account the cosmological expansion
yields the net change in the Stokes vector,
\beq
\De \mbf s = \int_z^0 \frac{i\Si_z\cdot\mbf s}{(1+z)H_z}\ dz\ ,
\label{dstokes}
\eeq
where $\Si_z$ is the rotation matrix 
at the blue-shifted frequency $(1+z)\om$ 
and source direction $\mbf{\hat n}$.
The net polarization change $\De \mbf s$ 
can depend on frequency and direction of propagation.
To seek birefringence,
we can either model the polarization
at the source and search for discrepancies 
in the observed polarization,
or we can test for unexpected frequency dependences.

In what follows,
we investigate vacuum birefringence via the CMB,
leaving the use of GRB polarimetry in this context 
to be discussed elsewhere 
\citep{km-long}.
The CMB has a long baseline but comparatively low frequency,
which implies lesser sensitivities to $d>3$ violations
relative to higher-frequency sources.
Here,
we focus on the four $d=3$ Lorentz-violating operators.
These induce energy-independent polarization changes,
so the best constraints are expected from the most distant sources
irrespective of frequency.
The CMB therefore has the potential to yield maximal sensitivity
to these CPT-odd operators.
For any CPT-odd case,
birefringence causes a rotation of the Stokes vector
about the $s^3$ axis,
corresponding to a rotation 
of the linear-polarization angle $\ps$
with no change in the degree of linear or circular polarization.
For $d=3$,
the value of $\ps$ at present is 
$\ps = \ps_z + \de\ps_z$,
where $\ps_z$ is the blueshifted angle
and $\de\ps_z$ is its rotation,
\beq
\de\psi_z=
\int_0^z \frac{dz} {(1+z)H_z}
\sum_{jm} \syjm{0}{jm}(\mbf{\hat n})\ \kVt .
\label{delta_psi}
\eeq
Taking $z=1100$ for the CMB
and including a small radiation component $\Om_r\simeq 0.015$,
the rotation reduces to the direction-dependent approximation
\beq
\de\psi_{\rm CMB} \simeq
3.5^\circ\times10^{43}\ {\rm GeV}^{-1}
\sum_{jm}
\syjm{0}{jm}(\mbf{\hat n})\
\kVdjm{3}{jm} \ .  
\label{cmb_delta_psi}
\eeq
We remark in passing that CPT-even operators
produce a complicated mixing of circular and linear polarization,
rather than a simple rotation of $\psi$
\citep{km-long}.

We next search for the above effect in the
recent WMAP 5-year results
\citep{wmap5yr1,wmap5yr2}.
We generate initial sky maps of the Stokes parameters
using the best fit correlation coefficients $C_j$
as calculated by the WMAP collaboration
within the $\La$-CDM model
assuming gaussianality.
The Stokes parameters at each point on the sky 
are then rotated appropriately and used 
to calculate the $C_j$ coefficients
at the present epoch.
The likelihood of these coefficients 
is determined using available WMAP software.
The underlying cosmology is kept fixed,
so we are comparing the likelihood of Lorentz violation 
relative to a reasonable Lorentz-invariant cosmology.
Our analysis uses $TE$ and $TB$ data at high-$l$ 
corresponding to $j=24$-450,
disregarding the $TT$ data.
The latter is a good approximation because
the $TT$ data would dominate 
an analysis with varying cosmology
and therefore hold the cosmology comparatively fixed.

\begin{figure}
\epsscale{1}
\plotone{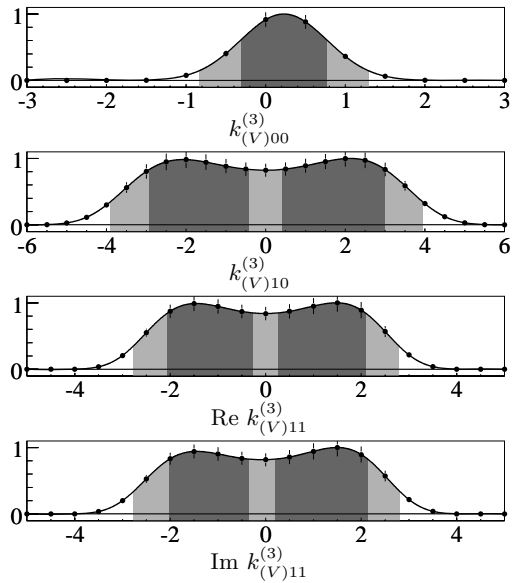}
\caption{\label{like}
Relative likelihood versus the four $d=3$ coefficients
for Lorentz violation.
Points indicate the values at which the ensemble averages
were made,
and the bars represent their standard errors.
The line is an extrapolation through the points.
Dark-gray regions indicate the estimated 68\% confidence interval,
while the light-gray region shows the 95\% level.
All coefficients are in units of $10^{-42}$ GeV.}
\end{figure}

The correlation coefficients $C_j$ are rotationally invariant,
so our analysis has sensitivity only 
to rotationally invariant combinations
of Lorentz-violating coefficients.
In the present context,
these are the isotropic coefficient 
$(\kafd{3})^T=- \kVdjm{3}{00}/\sqrt{4\pi}$
and the scalar magnitude 
$|\mbf{\kafd{3}}|=
\big(6|\kVdjm{3}{11}|^2+3|\kVdjm{3}{10}|^2\big)^{1/2}/\sqrt{4\pi}$.
In particular,
our results are independent of the direction of $\mbf{\kafd{3}}$.
Although the analysis contains no {\it a priori} anisotropies,
the procedure involves generating random realizations 
that contain anisotropies.
As a result, the likelihood $L\big(\kVdjm{3}{jm},r\big)$
for a given realization $r$ is anisotropic.
In obtaining the total likelihood 
for a given set of coefficients for Lorentz violation,
we sum over the likelihoods of all possible realizations
weighted by the probability density $P(r)$,
yielding
$L\big(\kVdjm{3}{jm}\big) =\sum_r P(r) L\big(\kVdjm{3}{jm},r\big)$.
This total likelihood 
is simply the average over all possible universes
and is a rotationally invariant indicator of Lorentz violation.
Here,
we estimate 
$L\big(\kVdjm{3}{jm}\big)$ for a range of 
values of $\kVdjm{3}{jm}$
by averaging over 3,000 realizations per value.
The results for the four coefficients with $d=3$
are shown in Fig.\ \ref{like}.

For the isotropic coefficient,
we obtain the $1\si$ result 
\beq
\kVdjm{3}{00}=(2.3\pm5.4)\times 10^{-43} {\rm ~GeV}.
\label{iso}
\eeq
This improves by about an order of magnitude
on the previous limit from radiogalaxy data
\citep{cfj}.
The result is also consistent with that of the WMAP collaboration,
which found a rotation 
of $\de\psi \simeq 1.2^\circ \pm 2.2^\circ$ 
\citep{wmap_cpt}
corresponding to
$\kVdjm{3}{00} < (1.2\pm2.2)\times 10^{-43}$ GeV,
and that of another recent analysis yielding
$\kVdjm{3}{00} < 2.5\times 10^{-43}$ GeV 
\citep{kdm}.
Other reported limits obtained from WMAP 3-year data 
\citep{wmap3yr}
include
$\kVdjm{3}{00} = (6.0\pm 4.0) \times 10^{-43}$ GeV 
\citep{cmb_bire12}
and
$\kVdjm{3}{00} = (2.5\pm 3.0) \times 10^{-43}$ GeV 
\citep{cmb_bire13}.
Some indication of a nonzero rotation has been found
in previous studies.
One involving BOOMERANG (B03) data 
\citep{boomerang1,boomerang2}
alone yielded the possibility 
$\kVdjm{3}{00} = (12\pm 7) \times 10^{-43}$ GeV 
\citep{km_cmb},
while the result from another analysis 
combining B03 and WMAP 5-year data 
corresponds to
$\kVdjm{3}{00} < (2.6\pm1.9)\times 10^{-43}$ GeV
\citep{xia}.
While consistent with the latter,
our result \rf{iso}
shows little evidence for isotropic Lorentz violation.

For each anisotropic coefficient,
Fig.\ \ref{like} displays the likelihood.
As expected, 
the results are independent
of the direction of $\mbf{\kafd{3}}$.
The plot symmetries reflect the expected behavior under
$\mbf{\kafd{3}}\rightarrow -\mbf{\kafd{3}}$.
These plots yield the result
\beq
|\mbf{\kafd{3}}|=(10^{+4}_{-8})\times 10^{-43} {\rm ~GeV},
\label{aniso}
\eeq
revealing some evidence at 1$\si$ 
for anisotropic Lorentz violation in the WMAP 5-year data.
This agrees with the indication 
of anisotropic Lorentz violation found 
from an analysis of B03 data,
which corresponds to 
$|\mbf{\kafd{3}}|=(15\pm 6)\times 10^{-43}$ GeV
\citep{km_cmb}.
The data are consistent with no Lorentz violation at $2\si$,
with a 95\% confidence level of 
$|\mbf{\kafd{3}}|<2\times 10^{-42}$ GeV.
This fully constrains the vector components 
of $\mbf{\kafd{3}}$,
and the results \rf{iso} and \rf{aniso}
provide a measurement of all four of the $d=3$
coefficients for Lorentz violation.

\acknowledgments
This work is supported in part by the U.S.\ D.o.E.\
under grant DE-FG02-91ER40661.


\begin{thebibliography}{}

\bibitem[Albert \etal (2008)]{magic}
Albert,
J.\ 
\etal\
2008,
Phys.\ Lett.\ B {\bf 668}, 253 

\bibitem[Amelino-Camelia \etal (1998)]{disp}
Amelino-Camelia,
G.,
Ellis,
J.,
Mavromatos,
N.E.,
Nanopoulos,
D.V.,
and 
Sarkar, 
S.\ 
1998,
Nature {\bf 393}, 763 

\bibitem[Amelino-Camelia \& Piran(2001)]{threshold}
Amelino-Camelia,
G.,
and
Piran, T.\ 
Phys.\ Rev.\ D {\bf 64}, 036005 

\bibitem[Amelino-Camelia(2008)]{review}
Amelino-Camelia,
G.\
2008, 
arXiv:0806.0339

\bibitem[Antonini \etal (2005)]{cavities2}
Antonini, 
P.\ 
\etal\
2005,
Phys.\ Rev.\ A {\bf  71}, 050101(R)

\bibitem[Boggs \etal (2004)]{boggs}
Boggs, 
S.E.,
Wunderer,
C.B.,
Hurley,
K.,
and
Coburn,
W.\ 
2004,
Astrophys.\ J.\ {\bf 611}, L77 

\bibitem[Cabella \etal (2007)]{cmb_bire13}
Cabella P., 
Natoli P.,
and 
Silk, J.\ 
2007,
Phys.\ Rev.\ D {\bf 76}, 123014 

\bibitem[Carroll \etal (1990)]{cfj}
Carroll,
S.M.,
Field,
G.B.,
and
Jackiw,
R.\ 
1990,
Phys.\ Rev.\ D {\bf 41}, 1231 

\bibitem[Colladay \& Kosteleck\'y(1997)]{sme1} 
Colladay, 
D.\
and 
Kosteleck\'y,
V.A.\ 
1997,
Phys.\ Rev.\ D {\bf 55}, 6760 

\bibitem[Colladay \& Kosteleck\'y(1998)]{sme2} 
Colladay,
D.\
and 
Kosteleck\'y,
V.A.\ 
1998,
Phys.\ Rev.\ D {\bf 58}, 116002

\bibitem[Ellis \etal (2006)]{emnss}
Ellis,
J.,
Mavromatos,
N.E.,
Nanopoulos, 
D.V.,
Sakharov,
A.S.,
and 
Sarkisyan,
E.K.G.\ 
2006,
Astropart.\ Phys.\ {\bf 25} 402

\bibitem[Fan \etal (2007)]{grb_bire4}
Fan, Y.-Z., 
Wei, D.-M.,
and Xu, D.\ 
2007, 
Mon.\ Not.\ R.\ Astron.\ Soc.\ {\bf 376}, 1857 

\bibitem[Feng \etal (2006)]{cmb_bire12}
Feng, 
B., 
Li,
M.,
Xia,
J.-Q.,
Chen,
X.,
and
Zhang, 
X.\
2006,
Phys.\ Rev.\ Lett.\ {\bf 96}, 221302 

\bibitem[Gamboa \etal (2006)]{cmb_bire23}
Gamboa, J., 
L\'opez-Sarri\'on, J., 
and 
Polychronakos, A.P.\ 
2006,
Phys.\ Lett.\ B {\bf 634},  471 

\bibitem[Goldberg(1967)]{sYjm2}
Goldberg, 
J.N.\ 
1967,
J.\ Math.\ Phys.\ {\bf 8}, 2155 
  
\bibitem[Hinshaw \etal (2009)]{wmap5yr1}
Hinshaw, 
G.\ 
\etal\
2009,
Astrophys.\ J.\ Suppl., in press
[arXiv:0803.0732]

\bibitem[Jacob \& Piran(2008)]{jacob}
Jacob, U.\ 
and 
Piran, T.\ 
2008,
J.\ Cosmol.\ Astropart.\ Phys.\ {\bf 0801}, 031 

\bibitem[Jacobson \etal (2004)]{grb_bire2}
Jacobson, T.,
Liberati, S., 
Mattingly, D.,
and 
Stecker, F.W.\
2004, 
Phys.\ Rev.\ Lett.\ {\bf 93}, 021101 

\bibitem[Kahniashvili \etal (2006)]{grb_bire3}
Kahniashvili, 
T.\ 
Gogoberidze, 
G.\ 
and 
Ratra,
B.\ 
2006,
Phys.\ Lett.\ B {\bf 643}, 81 

\bibitem[Kahniashvili \etal (2008)]{kdm}
Kahniashvili, 
T., 
Durrer, 
R., 
and 
Maravin, 
Y.\
2008, 
arXiv:0807.2593

\bibitem[Komatsu \etal (2009)]{wmap_cpt}
Komatsu,
E.\ 
\etal\
2009,
Astrophys.\ J.\ Suppl., in press
[arXiv:0803.0547]

\bibitem[Kosteleck\'y(2004)]{sme3} 
Kosteleck\'y,
V.A.\ 
2004,
Phys.\ Rev.\ D {\bf 69}, 105009 
  
\bibitem[Kosteleck\'y \& Mewes(2001)]{km_agn}
Kosteleck\'y,
V.A.\ 
and 
Mewes,
M.\ 
2001,
Phys.\ Rev.\ Lett.\ {\bf 87}, 251304 

\bibitem[Kosteleck\'y \& Mewes(2002)]{km}
Kosteleck\'y, 
V.A.\ 
and 
Mewes,
M.\ 
2002,
Phys.\ Rev.\ D {\bf 66}, 056005 

\bibitem[Kosteleck\'y \& Mewes(2006)]{km_grb}
Kosteleck\'y, 
V.A.\ 
and 
Mewes,
M.\ 
2006,
Phys.\ Rev.\ Lett.\ {\bf 97}, 140401

\bibitem[Kosteleck\'y \& Mewes(2007)]{km_cmb}
Kosteleck\'y, 
V.A.\ 
and 
Mewes,
M.\ 
2007,
Phys.\ Rev.\ Lett.\ {\bf 99}, 011601 

\bibitem[Kosteleck\'y \& Mewes(2009)]{km-long}
Kosteleck\'y, 
V.A.\ 
and 
Mewes,
M.\ 
2009,
arXiv:0905.0031
 
\bibitem[Kosteleck\'y \& Potting(1991)]{kp91}
Kosteleck\'y, 
V.A.\ 
and 
Potting,
R.\
1991,
Nucl.\ Phys.\ B {\bf 359}, 545

\bibitem[Kosteleck\'y \& Potting(1995)]{kp95}
Kosteleck\'y, 
V.A.\ 
and 
Potting,
R.\
1995,
Phys.\ Rev.\ D {\bf 51}, 3923 

\bibitem[Kosteleck\'y \& Russell(2008)]{tables}
Kosteleck\'y, 
V.A.\ 
and 
Russell,
N.\ 
2008, 
{\it Data Tables for Lorentz and CPT Violation,}
arXiv:0801.0287

\bibitem[Kosteleck\'y \& Samuel(1989)]{ks89}
Kosteleck\'y, 
V.A.\ 
and 
Samuel,
S.\
1989,
Phys.\ Rev.\ D {\bf 39}, 683

\bibitem[Lamon \etal (2008)]{lps}
Lamon,
R.,
Produit,
N.,
and  
Steiner,
F.\
2008,
Gen.\ Rel.\ Grav.\ {\bf 40}, 1731 

\bibitem[Lipa \etal (2003)]{cavities1}
Lipa,
J.\ 
\etal\
2003,
Phys.\ Rev.\ Lett.\ {\bf  90}, 060403 

\bibitem[Mart\'\i nez \& Piran(2006)]{mmtp}
Mart\'\i nez, 
M.R.\
and
Piran, 
T.\
2006,
JCAP 0604:006

\bibitem[Mitrofanov(2003)]{grb_bire1}
Mitrofanov, 
I.G.\ 
2003,
Nature {\bf 426}, 139 

\bibitem[Montroy \etal (2006)]{boomerang1}
Montroy,
T.E.\ 
\etal\ 
2006,
Astrophys.\ J.\ {\bf 647}, 813 

\bibitem[M\"uller \etal (2007)]{cavities4}
M\"uller,
H.\
\etal\
2007,
Phys.\ Rev.\ Lett.\ 
{\bf 99}, 050401

\bibitem[Newman \& Penrose(1966)]{sYjm1}
Newman,
E.T.\ 
and
Penrose, 
R.\ 
1966,
J.\ Math.\ Phys.\ {\bf 7}, 863 

\bibitem[Nolta \etal (2009)]{wmap5yr2}
Nolta,
M.\ 
\etal\
2009,
Astrophys.\ J.\ Suppl., in press
[arXiv:0803.0593]

\bibitem[Page \etal (2007)]{wmap3yr}
Page,
L.\ 
\etal\
2007,
Astrophys.\ J.\ Supp.\ {\bf 170}, 335

\bibitem[Piacentini \etal (2006)]{boomerang2}
Piacentini,
F.\ 
\etal\ 
2006,
Astrophys.\ J.\ {\bf 647}, 833 

\bibitem[Xia \etal (2008)]{xia}
Xia,
J.-Q.,
Li,
H.,
Zhao,
G.-B.,
and
Zhang,
X.\ 
2008,
Astrophys.\ J.\ {\bf 679}, L61 

\end{thebibliography}
\end{document}